\documentclass{elsart}
 \usepackage{graphics}
 \usepackage{graphicx}
 \usepackage{epsfig}
 \usepackage{amssymb}

\begin{document}

\begin{frontmatter}

\title{\boldmath A Study of $J/\psi\to \gamma\gamma V (\rho, \phi)$ 
      Decays with the BESII Detector}

\begin{small}
\begin{center}

\vspace{0.2cm}

J.~Z.~Bai$^1$,        Y.~Ban$^{10}$,         J.~G.~Bian$^1$,
X.~Cai$^{1}$,          J.~F.~Chang$^1$,
H.~F.~Chen$^{15}$,    H.~S.~Chen$^1$,        H.~X.~Chen$^{1}$,
J.~C.~Chen$^1$,        Jin~Chen$^{1}$
% ...
\\(BES Collaboration)\\ 

\vspace{0.2cm}
\label{att}
$^1$ Institute of High Energy Physics, Beijing 100039, People's Republic of
     China\\
$^2$ China Center of Advanced Science and Technology, Beijing 100080,
     People's Republic of China\\

$^3$ Guangxi Normal University, Guilin 541004, People's Republic of China\\

$^4$ Henan Normal University, Xinxiang 453002, People's Republic of China\\
$^5$ Huazhong Normal University, Wuhan 430079, People's Republic of China\\
$^6$ Hunan University, Changsha 410082, People's Republic of China\\
                                                    
$^7$ Liaoning University, Shenyang 110036, People's Republic of China\\

$^{8}$ Nanjing Normal University, Nanjing 210097, People's Republic of China\\

$^{9}$ Nankai University, Tianjin 300071, People's Republic of China\\
$^{10}$ Peking University, Beijing 100871, People's Republic of China\\
$^{11}$ Shandong University, Jinan 250100, People's Republic of China\\
$^{12}$ Shanghai Jiaotong University, Shanghai 200030, 
        People's Republic of China\\
$^{13}$ Sichuan University, Chengdu 610064,
        People's Republic of China\\                                    
$^{14}$ Tsinghua University, Beijing 100084, 
        People's Republic of China\\
$^{15}$ University of Science and Technology of China, Hefei 230026,
        People's Republic of China\\
$^{16}$ Wuhan University, Wuhan 430072, People's Republic of China\\
$^{17}$ Zhejiang University, Hangzhou 310028, People's Republic of China\\
\vspace{0.4cm}

$^{\ast}$ Visiting professor to University of Michigan, Ann Arbor, MI 48109 USA 
\end{center}
\end{small}
\vspace{0.4cm}

\begin{abstract}
       Using a sample of $58\times 10^6$ $J/\psi$ events collected
       with the BESII detector, radiative decays
       $J/\psi\rightarrow\gamma\gamma V$, where $V=\rho$ or $\phi$,
       are studied. A resonance around 1420 MeV/c$^2$ (X(1424)) is
       observed in the $\gamma\rho$ mass spectrum. Its mass and width
       are measured to be $1424\pm 10 \:({\rm stat})\pm 11 \:({\rm
       sys})$ MeV/c$^2$ and $ 101.0\pm 8.8 \pm 8.8$
       MeV/c$^2$, respectively, and its branching ratio $B(J/\psi\to \gamma
       X(1424)\to \gamma\gamma \rho)$ is determined to be
       $(1.07\pm0.17 \pm 0.11)\times 10^{-4}$.
       A search for $X(1424)\to \gamma\phi$ yields a 95\%~C.L. upper
       limit $B(J/\psi\to \gamma X(1424)\to \gamma\gamma \phi) < 0.82
       \times 10^{-4}$.
\end{abstract}

\end{frontmatter}

   \section{Introduction}\label{int}

Experimentally the structure of the $\eta(1440)$ remains unresolved.
The existence of two overlapping pseudo-scalar states has been
suggested: one around 1410 MeV/c$^2$ decays into both $ K\overline{K} \pi$
and $\eta \pi\pi$, and the other around 1470 MeV/c$^2$ decays only to $ K
\overline{K} \pi$ \cite{pdg,godfrey}. It is therefore conceivable that
the higher mass state is the $s\bar{s}$ member of the $2^1S_0$ nonet
\cite{rath}, while the lower mass state may contain a large gluonic
content \cite{close}.

Standard perturbative theory predicts \cite{chan} that if the
$\eta(1440)$ is a $q \overline q$ state which decays in a flavor
independent way, the partial width relationship between its
$\gamma\rho$, and $\gamma\phi$ final states should be
$\Gamma_{\gamma\rho} : \Gamma_{\gamma\phi} $ = $ 9:2 $.
A simultaneous search for a
resonance near 1440 MeV/c$^2$ in the $\gamma\rho$ and $\gamma\phi$ mass
spectra and a determination of the branching ratios of the resonance
may shed light on the internal structure of the $\eta(1440)$.

Radiative decays of a resonance near 1440 MeV/c$^2$ to $\gamma V(V= \rho$,
and $\phi)$ have been studied previously in
$J/\psi\rightarrow\gamma\gamma V$ events by Crystal-Ball \cite{ewd},
MarkIII \cite{cof} and DM2 \cite{aug}.  The situation here is further
complicated by the proximity of the $f_1(1420)$ to the
$\eta(1440)$. MarkIII finds that the $0^-$ is only slightly favored
over the $1^+$ in a fit to their angular distributions in
$J/\psi\rightarrow\gamma\gamma \rho$ \cite{cof}.

In this letter, we report a study of decays $J/\psi\to \gamma \gamma
      \rho $ and $J/\psi\to \gamma \gamma \phi $ selected from a
      sample of $58\times 10^6$ $J/\psi$ events collected by the
      Beijing Spectrometer (BESII)
      detector.

\section{Event selection}
      We want to study
\begin{center}
           $J/\psi\rightarrow\gamma X$, $X\rightarrow\gamma V,$ 
\end{center}
     where $X$ is a resonance, and $V$ denotes vector mesons $\rho$ or $\phi$,
     which are reconstructed via their decays
    $ \rho\rightarrow\pi^+\pi^-$ and 
    $ \phi\rightarrow K^+ K^-$.

The BESII detector has been described in detail elsewhere \cite{bai}.
In this study two oppositely charged particles must be detected in the
main drift chamber.  Photons are detected by the barrel
shower counter (BSC) which covers $80\% $ of the $4\pi$ solid angle
with an energy resolution $\delta E/E = 21 \% /\sqrt{E}$.  In order to
remove electronic noise, the energy deposited in the BSC by each neutral
particle is required to have a minimum of 70 MeV.  A photon is
required to be isolated from charged tracks (cos $\theta _{\gamma
\pi(K)}<0.98 $, where $\theta_{\gamma \pi(K)} $ is the angle between
the photon and a charged particle) to reject any photons radiated by a
charged particle in the event, and to be consistent with
originating from the event interaction point.  Photon candidates
satisfying these criteria are used for this analysis.  The highest
energy photon in an event is taken as the radiative photon directly
produced in $J/\psi\to \gamma X$ events.

\subsection{\boldmath\bf Selection of 
        $J/\psi\to \gamma\gamma\rho \to \gamma\gamma \pi^+ \pi^-$ events}\label{sec}
Monte Carlo (MC) simulations have been carried out for both
the signal and background processes. The backgrounds considered here
are radiative $J/\psi$ decays into two charged tracks, namely
$(m\gamma)\pi^+\pi^-$ (m=1,2,3,4) and $(n\gamma)K^+K^-$ (n=1,2,3,4)
for which known branching ratios compiled by the Particle Data Group
(PDG)\cite{pdg} are used to form the correct mixture of these
processes in the Monte Carlo background simulation.  The most
important backgrounds for this channel are $J/\psi\to \gamma \eta_c$,
$J/\psi\to \gamma \eta \pi \pi$, $J/\psi\to \gamma f_1(1510)\to \gamma
\eta \pi \pi$, $J/\psi\to\omega \pi^0$, $J/\psi\to\omega \eta$,
$J/\psi\to a_2(1320) \rho$, and $J/\psi\to b_1^0(1235)\pi^0$. The
generated Monte Carlo samples of signal and background are analyzed,
and selection variables are varied until an optimized ratio of signal to
background is reached. As a result,
the following criteria are chosen for the $J/\psi\to
\gamma\gamma \rho \to \gamma\gamma \pi^+ \pi^-$ analysis:

\begin{enumerate}
\item 
 The sum of momenta of the charged tracks in the event 
       ( $P_{miss}$) is less than 1.14 GeV/c,
\item
at least one of the two charged tracks in the event must have a higher
particle identification confidence level for the pion hypothesis than
for the kaon hypothesis by combining the information from TOF and dE/dx,
\item
 the $\chi^2$ of a four constraint kinematic fit of the event to a
 $\gamma\gamma \pi^+\pi^-$ topology is less than 10.0,
\item
the total energy of any photons not used in the kinematic fit in criterion
(3) is less than 250 MeV,
\item
the invariant mass of the two selected photons must be greater than 
0.66 GeV/c$^2$, and
\item
the helicity angle $\theta$ of the dipion in the $\gamma\pi^+\pi^-$ system
must satisfy $|\cos\theta|<0.86$.

\end{enumerate}

\begin{figure}[htb]
\begin{minipage}{70mm}
\begin{center}
\epsfig{file=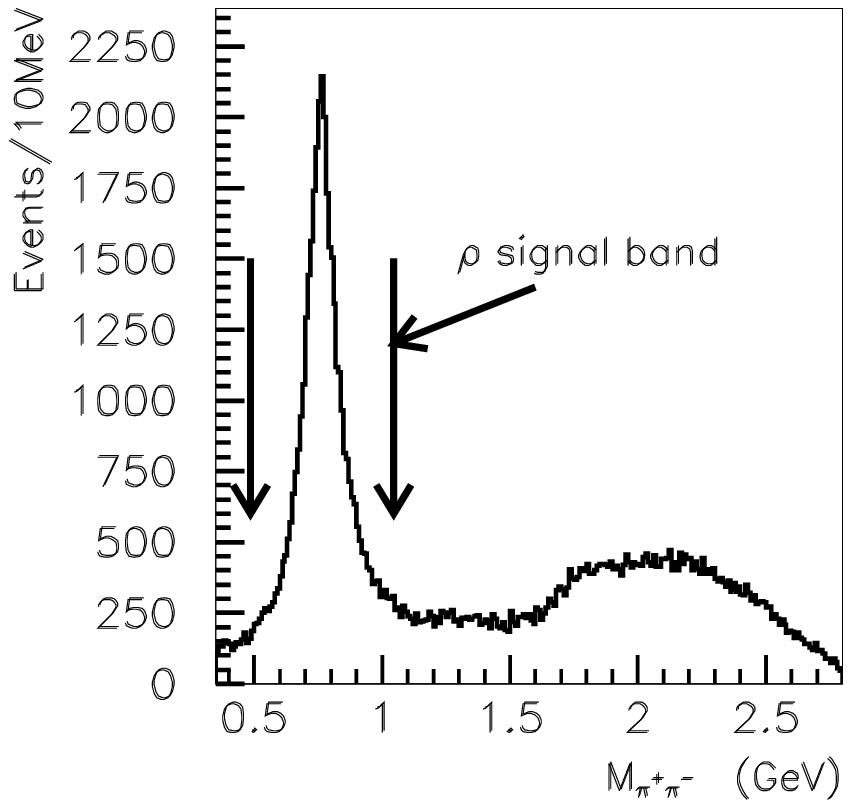 ,bbllx=33pt,bblly=417pt,%
         bburx=276pt,bbury=665pt,width=6.5cm,height=7cm,clip=}
\caption{The $\pi^+\pi^-$ invariant mass distribution.}
\end{center}
\label{fig1}
\end{minipage}
\hspace{\fill}
\begin{minipage}{70mm}
\begin{center}
\epsfig{file=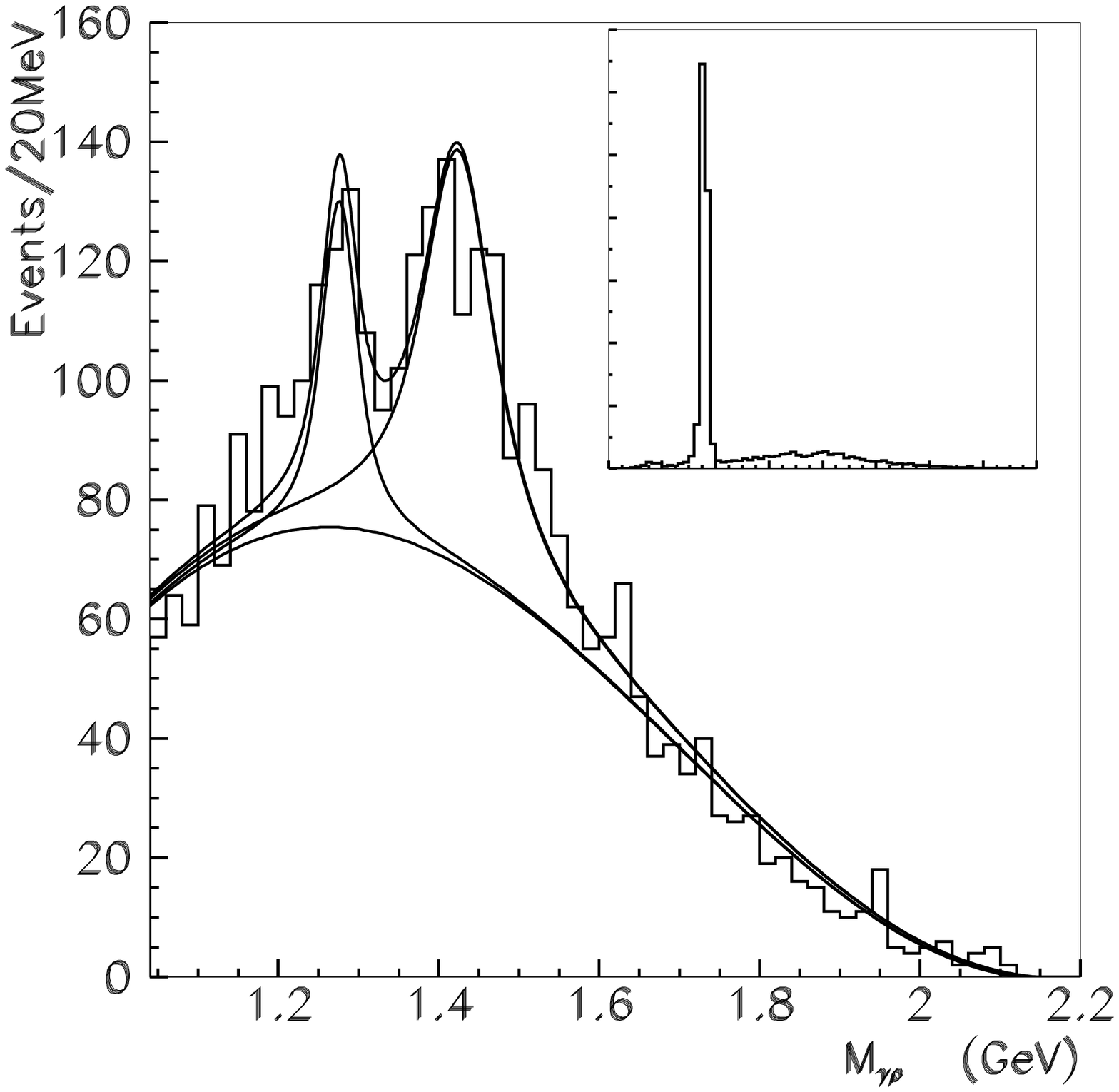 ,bbllx=57pt,bblly=164pt,%
         bburx=534pt,bbury=648pt,width=5cm,height=5cm,clip=}  
\end{center}
\caption{The $\gamma\rho$ invariant mass distribution. The insert
shows the full mass scale where the $\eta(958)$ is clearly observed. }
\label{fig2}
\end{minipage}
\end{figure}

Fig. 1 shows the $\pi^+\pi^-$ invariant mass spectrum
 of the selected
$\gamma\gamma\pi^+\pi^-$ events, where a clear $\rho$ signal is
visible.  To select $\rho$ candidates, $\pi^+\pi^-$ pairs must satisfy
$|M_{\pi^+\pi^-}-M_\rho| \leq 0.28$ GeV/c$^2$, as indicated in
Fig. 1.  Combining the $\rho$ candidate with the lower
energy photon in the $\gamma\gamma\pi^+\pi^-$ event, the $\gamma
\rho$ mass distribution,
shown in Fig.~\ref{fig2}, is obtained.
\par
\par
\subsection{\boldmath\bf Selection of 
    $J/\psi\to \gamma\gamma\phi \to  \gamma \gamma K^+ K^-$ events}\label{sec1}
The selection criteria for $J/\psi\to \gamma X \to \gamma\gamma\phi
~(\phi \rightarrow K^+ K^-)$ have been chosen in a similar way to
those for the $\gamma\gamma\pi^+\pi^-$.  According to the Monte Carlo
simulation, we find that the major sources of backgrounds are from
$J/\psi\to \phi f_0(980)$, $J/\psi\to \gamma \eta(1440) \to \gamma K
\bar{K} \pi$, and $J/\psi\to \gamma f_1(1420) \to \gamma K \bar{K}
\pi$.  The final requirements are: $P_{miss}\leq 1.31 $ GeV/c,
$\chi^2 \leq 25$ for the four constraint kinematic fit to the
$J/\psi\to \gamma\gamma K^+K^-$ hypothesis, and at least one
identified charged kaon must be present in the event. The other criteria
remain the same for this mode as for the $\gamma\gamma\pi^+\pi^-$
analysis.
\par
\begin{figure}[htb]
\begin{minipage}{70mm}
\begin{center}
\epsfig{file=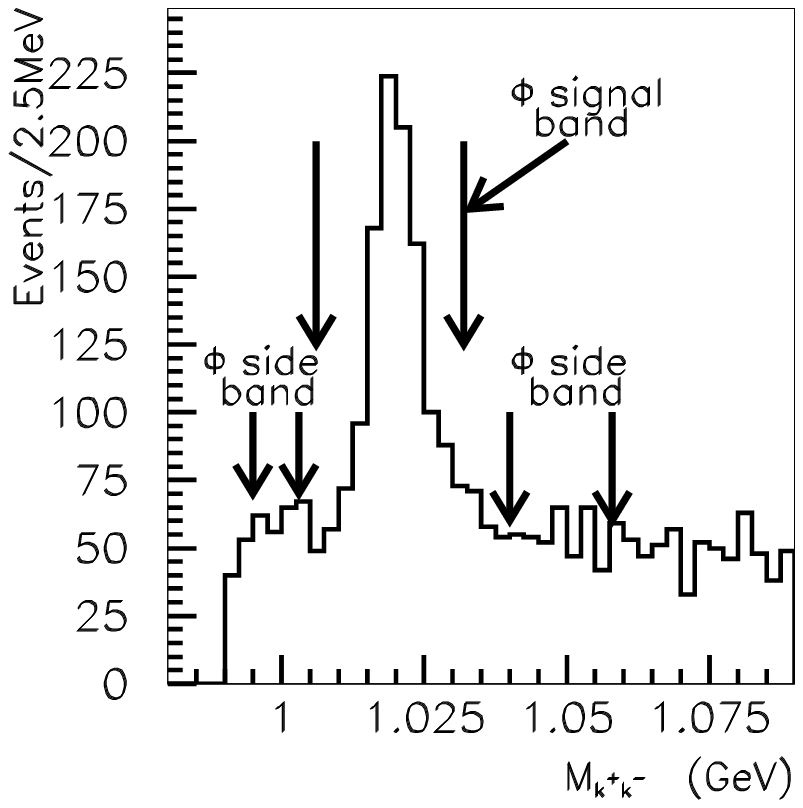 ,bbllx=50pt,bblly=412pt,%
         bburx=288pt,bbury=665pt,width=6.5cm,height=7cm,clip=}
\caption{The  $K^+ K^-$ invariant mass distribution.}
\label{fig3}
\end{center}
\end{minipage}
\hspace{\fill}
\begin{minipage}{70mm}
\begin{center}
\epsfig{file=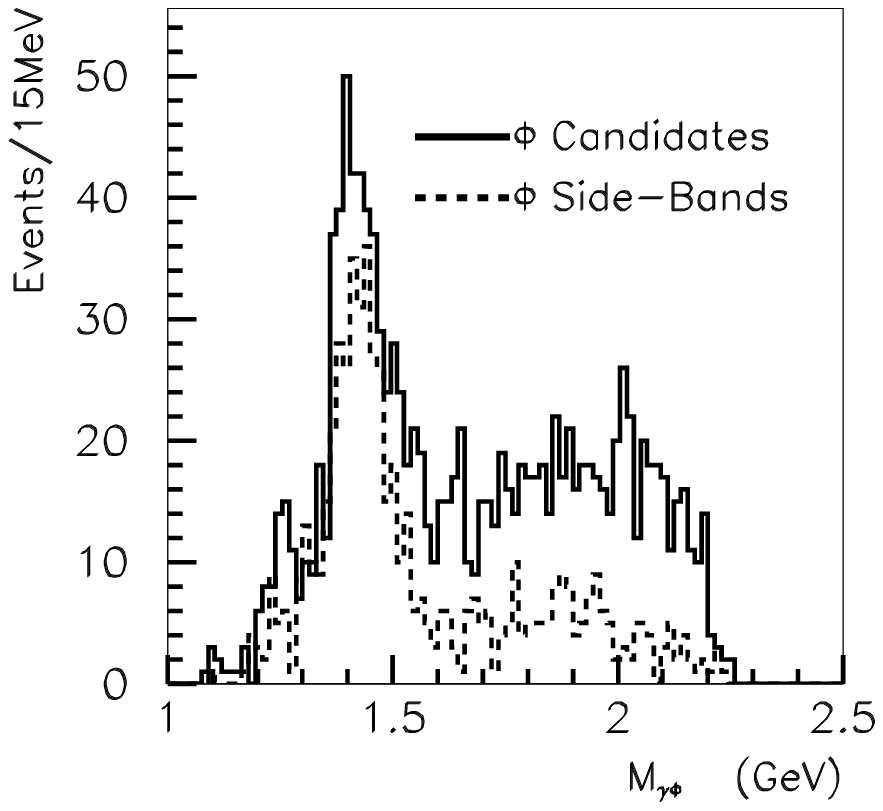 ,bbllx=41pt,bblly=421pt,%
         bburx=288pt,bbury=669pt,width=6.5cm,height=7cm,clip=}
\caption{The  $\gamma \phi$ invariant mass distribution. }
\label{fig4}
\end{center}
\end{minipage}
\end{figure}
\par
Fig.~\ref{fig3} shows the $K^+K^-$ mass distribution of the selected events.
The $\phi$ candidates must satisfy
$|M_{K^+ K^-}-M_\phi| \le 0.013$ GeV/c$^2$. Also
shown in Fig.~\ref{fig3} are the side-band regions, as well as the $\phi$
signal region. The
$\gamma\phi$ mass distribution is shown in Fig.~\ref{fig4}. The lower energy
$\gamma$ is also combined with the $\phi$ side-band events forming the
dashed distribution, also shown in Fig.~\ref{fig4}.

\section{\boldmath\bf  Analysis and Results}
\subsection{\boldmath\bf  $J/\psi\to \gamma\gamma\rho \to \gamma\gamma
  \pi^+ \pi^-$ events} 
We have fitted the mass distributions in Figs. 1 and \ref{fig3}, and estimate
that they contain 38249$\pm$490 $\rho$ and 764$\pm$64 $\phi$ events,
respectively.  The insert in Fig.~\ref{fig2} shows the full
$\gamma\rho$ mass range, where a strong $J/\psi\to \gamma
\eta^\prime(958)\to \gamma\gamma\pi^+\pi^-$ signal is observed, as
expected. To verify that the mass scale is correct, we have fitted the
$\eta(958)$ signal and obtain 957.5$\pm$0.2 MeV/c$^2$ for the mass and
0.20$\pm$0.04 MeV/c$^2$ for the width, which are in excellent agreement with
the world average values of 957.78$\pm$0.14 MeV/c$^2$ and 0.202$\pm$0.016
MeV/c$^2$, respectively.

Two enhancements above 1.2 GeV/c$^2$ in the $\gamma\rho$ mass spectrum
are evident in Fig.~\ref{fig2}.  We have examined the $\gamma \pi^+\pi^-$ mass
distribution for $\pi^+\pi^-$ pairs with masses just above the upper
edge of the $\rho$ mass band. The distribution does not exhibit any
distinct structures.  We conclude that the peaks in Fig.~\ref{fig2}
are associated with the $\rho$.

The identical selection criteria have been applied to a sample of 30 million
Monte Carlo inclusive $J/\psi$ events which do not contain the decay
$J/\psi\to \gamma X(1420)\to \gamma \gamma\rho\to \gamma\gamma\pi^+\pi^-$.  The
resulting Monte Carlo $\gamma\pi^+\pi^-$ mass distribution does not show the
enhancement at 1420 MeV/c$^2$ but does show the $f_1(1285)$, as
expected.

In order to extract the resonance parameters in Fig.~\ref{fig2}, we
perform an unbinned maximum-likelihood fit to the data. The fit
function consists of two Breit-Wigner functions, each convoluted with
a Gaussian with a mass resolution of 12 MeV/c$^2$, for the signals
($f_1(1285)$ and X(1424)), and a polynomial function for the
background.  The $\chi^2/$dof of the fit is $68.3/48$. In order to
check whether the background shape in our fit is correct, we compared
it with the background from our $J/\psi$ inclusive MC sample and find
that the backgrounds are consistent.
The results of the fit are shown in Fig.~\ref{fig2} and summarized in
Table~\ref{tab2}, where the first errors are statistical errors
obtained from the fit and the second are systematic.  

The systematic errors on the mass and the width for the first
resonance $(1276)$ are determined from the variations when different
background functions are used in the fit, about $0.07\%$ and
$19.5\%$, respectively, and from the uncertainty of the Monte Carlo
simulation, about $0.6\%$ and $12.5\%$, respectively.  The systematic
errors on the mass and the width for the second resonance $(1424)$
include the background function variations, about $0.01\%$ and
$2.2\%$, respectively, and the uncertainty of the Monte Carlo
simulation, about $0.8\%$ and $8.4\%$, respectively.

The detection efficiencies for $J/\psi\to \gamma X \to \gamma
\gamma\rho\to \gamma\gamma\pi^+\pi^-$ are determined from a Monte
Carlo simulation to be $(9.3\pm 0.1)\%$ at 1.285 GeV/c$^2$ and $(8.81\pm
0.09)\%$ at 1.420 GeV/c$^2$. The systematic errors on the
branching ratios are determined by combining the Monte Carlo
uncertainty on the efficiencies ($8.4\%$), the error on the number of
$J/\psi$ events ($5.0\%$), and the variation in the number of signal
events due to the different background shapes used in the fit ($13.2\%
$ and $6.5\% $ for the first and second resonances, respectively).

\par
\begin{table}[h]
\begin{center}
 \caption{$J/\psi\rightarrow\gamma X 
                (X\rightarrow\gamma\rho)$ results.}
  \begin{tabular}{|c|c|c|l|c|}\hline
     Mass  & Width & B$(J/\psi
     \rightarrow\gamma X \to \gamma\gamma\rho)$ & Events & Signi-\\
     (MeV/c$^2$) & (MeV/c$^2$) & $(\times 10^{-4})$ & & ficanc  \\ \hline

     $1276.1\pm 8.1\pm 8.0$ & $40.0\pm 8.6\pm 9.3 $ & $0.38\pm 0.09\pm 
                        0.06$ 
                        & $203\pm 49$  &$6.3 \sigma$ \\ \hline
     $1424\pm 10\pm 11$ & $101.0\pm 8.8\pm 8.8$ & $1.07\pm 0.17\pm 
                        0.11$ 
                        & $547 \pm 86$ & $9.3\sigma$ \\ \hline
\end{tabular}
\label{tab2}
\end{center}
\end{table}
\par

To determine whether the X(1424) is more likely to be the $f_1(1420)$ or
the $\eta(1440)$, 
we use the measurement \cite{wa102} by the WA102 collaboration 
   and PDG results to obtain

        $${B(J/\psi\to \gamma f_1(1420), f_1(1420)\to \gamma\rho)
             < 1.7 \times 10^{-5}} \:(95\%~C.L.) $$

    Comparing this limit to our measurement of  
    B$(J/\psi\to \gamma X(1424) \to \gamma \gamma \rho) = (1.07 \pm
    0.17\pm 0.11 ) \times 10^{-4}$, we conclude that of 
    $X(1424)$ in $J/\psi\to \gamma \gamma \rho^0$ channel 
    should be predominantly $\eta(1440)$. 
\par
\begin{table}[h]
\begin{center}
\caption{Comparison with other experiments}
\label{tab3}
\begin{tabular}{|c|c|c|c|c|} \hline
    Decay & Mass & Width
    & $B(J/\psi\to\gamma X) * $
    & Experi-  \\
    Mode  & (MeV/c$^2$) & (MeV/c$^2$) & $B( X \to\gamma V)$
    & ment \\ 
    & & & ($\times 10^{-4}$)& \\ \hline
    $ f_1(1285)$ 
    & $1281.9\pm 0.6$ & $24.0\pm 1.2$
    & 0.34 $\pm$ 0.09  & PDG~\cite{pdg} \\
   $\to\gamma\rho^0$ & 1271 $\pm$ 7 & 31 $\pm$ 14
    & 0.25 $\pm$ 0.07 $\pm$ 0.03 & MarkIII~\cite{cof} \\
    & 1276.1 $\pm$ 8.1 $\pm$ 8.0 & 40.0 $\pm$ 8.6 $\pm$ 9.3
    & 0.38 $\pm$ 0.09 $\pm$ 0.06 & BESII \\
      \hline\hline
    $\eta(1440)$ 
    & 1400-1470  & 50-80
    & 0.64 $\pm$ 0.12 $\pm$ 0.07 & PDG~\cite{pdg} \\
    $\to\gamma\rho^0$ &1432 $\pm$ 8  & 90 $\pm$ 26 & 0.64 $\pm$ 0.12 $\pm$ 0.07
        &MarkIII~\cite{cof} \\
    &1424 $\pm$ 10 $\pm$ 11 & $101.0 \pm 8.8 \pm 8.8$ & 
      $1.07\pm 0.17 \pm 0.11$ 
           &BESII \\ \hline
    $\eta(1440)$  & & & $<$ 0.82 (95\%~C.L) &
             BESII\\ 
     $\to\gamma\phi$  & & &  & \\ \hline
\end{tabular}
\end{center}
\end{table}
\par
   For the resonance around 1276 MeV/c$^2$, MarkIII \cite{cof} finds that
   the $1^+$ hypothesis is preferred over the $0^-$ by about 4
   $\sigma$, leading to the conclusion that it is $f_1(1285)$.  The
   BESII results for the mass, width, and branching fraction of the
   lower mass peak are consistent with those of MarkIII, as shown in 
   Table~\ref{tab3}. 
   From MarkIII analysis, it is not distinguishable between $0^-$ and $1^+$
   hypothesis for the X(1432) state, but the $0^-$ slightly better 
   than $1^+$ by about 2$\sigma$ .

\subsection{\boldmath\bf $J/\psi\to \gamma\gamma\phi \to  \gamma \gamma
  K^+ K^-$ events} 
In Fig.~\ref{fig4}, the $\gamma\phi$ and $\gamma
K^+K^-$ ($\phi$ side-bands) distributions are quite similar. 
\par
By combining the information from TOF and dE/dx one can clearly distinguish
K from $\pi$ for charged kaon momenta between 200 MeV/c and 800 MeV/c. 
Therefore the contamination from $\pi$ misidentified as $K$
can be neglected.   
\par
\begin{figure}[h]
\begin{center}
\epsfig{file=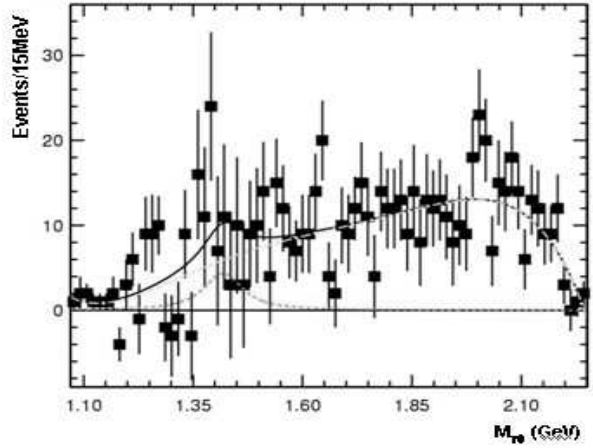,width=8cm,height=6cm,angle=0}
\caption{The invariant mass of $\gamma\phi$ after side-band 
                        background subtraction. }
\label{fig6}
\end{center}
\end{figure}

According to the Monte Carlo simulation, we find that the main backgrounds are 
as mentioned in Section~\ref{sec1}. They arise from the decays  $J/\psi\to 
\phi f_0(980)$, and $J/\psi\to \gamma \eta(1440) \to \gamma K \bar{K} \pi$ 
and $J/\psi\to \gamma f_1(1420)) \to \gamma K \bar{K} \pi$. These 
background events are very difficult to reject in our event selection.
For these processes a comparison of the two $\gamma K^+ K^-$ invariant mass 
spectra, derived from the $K^+ K^-$ mass around the $\phi$ signal and from 
$K^+ K^-$ from the $\phi$ side-bands, shows comparable contributions in the 
$\gamma K^+ K^-$ mass region around 1400 MeV/c$^2$. Therefore, the $\phi$ 
side-bands can be used to estimate the background in the $\gamma \phi$ 
spectrum.
In Fig.~\ref{fig6} the side-bands subtracted $\gamma\phi$ mass spectrum is 
shown, and no significant peak around 1420 MeV/c$^2$ is observed.
  
\par
\par

A $\chi^2$ fit is performed on the $\gamma\phi$ mass spectrum in
Fig.~\ref{fig6}. The fit function includes a Breit-Wigner function with mass
and width fixed to the values obtained in the fit to the peak at 1420
MeV/c$^2$ in Fig.~\ref{fig2}, convoluted with a Gaussian with a mass resolution
of 12 MeV/c$^2$ as predicted by the Monte Carlo simulation, and a
polynomial function for backgrounds.  The detection efficiency is
evaluated from a Monte Carlo simulation of the decay $J/\psi \to
\gamma \gamma \phi $ to be (5.4$\pm$0.1)\%.  We determine

     $$B(J/\psi \to \gamma X(1424)) B(X(1424)\to \gamma \phi) =
      (0.31\pm0.30)\times 10^{-4} $$

\noindent
which corresponds to a 95\% C. L. upper limit of

    $$B(J/\psi \to \gamma X(1424)) B(X(1424)\to \gamma \phi) <
      0.82\times 10^{-4} $$

   Table~\ref{tab3} shows a comparison of results from BESII (this 
work) and  other experiments.

\section{Conclusion}
\par
\noindent
If the X(1424) is the lower mass state of the 
$\eta(1440)$ as mentioned in \cite{pdg,godfrey}, it should be observed in 
the $J/\psi\to \gamma \gamma \phi$ channel. 
     Comparing our result on the branching ratio 
     $B(J/\psi \to \gamma X(1424) \to \gamma \gamma \rho) = (1.07 \pm
    0.17\pm 0.11 ) \times 10^{-4}$, and the 
upper limit of B($J/\psi \to \gamma X(1424) \to \gamma \gamma \phi)
   <0.82 \times 10^{-4}$ (95\% C.L.),  we cannot  draw a definite conclusion 
on wether the X(1424) is either a $q\bar{q}$ state or a glueball state. 
Therefore, further study is needed to clarify the situation.

\section{Acknowledgments}
 We would like to thank Profs. F.A.~Harris, X.C.~Lou, D.V. ~Bugg, H.~Yu 
and Dr. J.D.~Richman for valuable discussions. 
\par  
   The BES collaboration thanks the staff of BEPC for their hard efforts.
This work is supported in part by the National Natural Science Foundation
of China under contracts Nos. 19991480,10225524,10225525, the Chinese 
Academy
of Sciences under contract No. KJ 95T-03, the 100 Talents Program of CAS
under Contract Nos. U-11, U-24, U-25, and the Knowledge Innovation Project 
of
CAS under Contract Nos. U-602, U-34(IHEP); and by the National Natural Science
Foundation of China under Contract No.10175060(USTC).


\begin{thebibliography}{00}
\bibitem{pdg} K. Hagiwara {\it et al.}, Phys. Rev. {\bf D66} (2002) 010001
\bibitem{godfrey} S. Godfrey and J. Napolitano, Rev. Mod. Phys {\bf 71} (1999) 1411
\bibitem{rath}
              M. G. Rath {\it et al.}, Phys. Rev. D40 (1989) 693 ;\\
              A. Bertin {\it et al.}, Phys. Lett. B361 (1995) 187.
           \bibitem{close}  
              M. Acciarri et al., Phys. Lett. B501 (2001) 1 ;\\
              F. Close {\it et al.}, Phys. Rev. D55 (1997) 5749.
           \bibitem{chan} M. S. Chanowitz, Phys. Lett. B164 (1985) 379.
           \bibitem{ewd} C. Edwards, PhD thesis, Cal. Tech. Preprint
                      CALT-68-1165 (1985).
           \bibitem{cof} D. Coffman {\it et al.}, Phys. Rev. D41 (1990) 
                           1410;\\
                         J. D. Richman, PhD thesis, Caltech Preprint 
                            CALT-68-1231 (1985).
           \bibitem{aug}J. E. Augustin {\it et al.}, Orsay preprint  
                         LAL/85-27 (1985);\\
                        J. E. Augustin {\it et al.}, Phys. Rev. D42 
                           (1990) 10.
           \bibitem{bai}J. Z. Bai {\it et al.}, Nucl. Instr. Meth. Phys. 
                                   Res. A344(1994)319; \\
              J. Z. Bai {\it et al.}, Nucl. Instr. Meth. Phys.
                                   Res. A458(2001)627.
           \bibitem{wa102} D. Barberis {\it et al.}, Phys. Lett. B440 (1998) 225.
\end{thebibliography}
\end{document}